\documentclass[11pt,a4paper]{article}

\usepackage[english]{babel}% включение переносов

\usepackage{epigraph}
\usepackage{indentfirst}
\usepackage[unicode=true]{hyperref}
\usepackage{makeidx}
\makeindex
\usepackage{graphicx}
\usepackage[font=small,labelfont=bf]{caption}


\usepackage{latexsym}
\usepackage{amsfonts}
\usepackage{amsmath}
\usepackage{amssymb}

\font\twmsbm=msbm10 scaled 1200 \font\nmsbm=msbm9
\newfam\bbold
\textfont\bbold=\twmsbm \scriptfont\bbold=\nmsbm
\scriptscriptfont\bbold=\nmsbm

\font\twscr=rsfs10 scaled 1200 \font\nscr=rsfs10
\newfam\script
\textfont\script=\twscr \scriptfont\script=\nscr
\scriptscriptfont\script=\nscr

\newcommand{\be}{\begin{equation}}
\newcommand{\ee}{\end{equation}}
\newcommand{\bean}{\begin{eqnarray*}}
\newcommand{\eean}{\end{eqnarray*}}

\newcommand{\bea}{\begin{eqnarray}}
\newcommand{\eea}{\end{eqnarray}}

\author{M. G. Ivanov\thanks{e-mail: {\tt \href{email://mgi@mi.ras.ru}{mgi@mi.ras.ru}}}\\
{\small \emph{Laboratory of Quantum Information Theory,}}
{\small \emph{Moscow Institute of Physics and Technology}}\\
{\small \emph{Instituski per., 9, Dolgoprudnyi, Moscow region, Russia}}}
\title{On uniqueness of quantum measurement theory}
\date{August 23, 2015}
\begin{document}
\maketitle

\begin{abstract}
The paper discuss the structure of quantum mechanics and uniqueness of its postulates.

%The unitarity of evolution of closed quantum system is demonstrated to be a technical assumption, it have to be replaces by %isometricity.

The Born rule for quantum probabilities is fixed by requirement of nonexistence of quantum telepathy.

Von Neumann projection postulate describes the transformation of quantum state under the condition of no-interaction measurement. Projection postulate could be considered as transition to conditional probability under the condition of a certain result of quantum measurement.

\end{abstract}

\section{Introduction\label{sec-intro}: Quantum theory structure \label{sec-str-qt}}

In this section we present the decomposition of standard quantum theory into blocks, which have different status in contemporary science.

\begin{itemize}\itemsep=-1.5pt
\item \emph{The theory of closed quantum systems} is very well developed fundamental theory.
  It is time reversible, deterministic (no probabilities), stable with respect to initial data
  The entropy of system is conserved.
\item \emph{Measurement theory} is semiphenomenological theory of interaction of a system, which used to be closed, with
  a measuring device.
  It is time irreversible, probabilistic.
  The entropy of system+measuring device increases.
 \begin{itemize}\itemsep=-1.5pt
 \item \emph{Probability calculation for different measurement results (Born rule)} is
  fundamental rule.
 \item \emph{Calculation of quantum state after the measurement} is considered to be
 the most phenomenological part of quantum theory.
  Different approaches are suggested in literature.
  \begin{itemize}\itemsep=-1.5pt
   \item If the result of measurement is not (yet) known one has \emph{nonselective quantum measurement}.
   The theory of nonselective quantum measurement is phenomenological.
   It is well developed (decoherence), deterministic (no probabilities), irreversible.
   The entropy of system under nonselective measurement increases.
   \item If the result of measurement is known one has \emph{selective quantum measurement (wavefunction collapse)}.
   The theory of selective quantum measurement is phenomenological.
   It is considered to be the most disputable part of quantum theory.
   The major part of discussions on quantum theory interpretations and speculations is related with wavefunction collapse.
   It is probabilistic and irreversible.
   The entropy of system under selective measurement decreases.
  \end{itemize}
 \end{itemize}
\end{itemize}

   Any selective measurement have to be considered nonselective before one knows its result.
   So, quantum selective measurement could be splitted into two stages\\
   1) \emph{quantum nonselective measurement},\\
   2) \emph{``classical'' selective measurement} (the transition of \emph{classical} information from measuring device to an observer).\\

   The second stage is called ``classical'', but it is more miraculous.

   Quantum nonselective measurement nullify non-diagonal elements of density matrix in measurement basis.
   Classical selective measurement nullify all diagonal elements of density matrix except of a random one.
   The classical selective measurement is described by classical probability theory, so the choosing of one alternative in classical probability theory seems to be a trivial problem.
   Actually the problem is beyond the classical probability theory, and beyond the quantum theory.
   Theory of quantum decoherence describes the conversion of the superposition of pure states into a mix, but could not describe the choosing of the only alternative. So, the classical probability theory describes the probability, but could not describe the choosing of the only alternative too.

   Below we consider the exact measurement of observables with discreet spectra.

%\section{Unitarity or isometricity? \label{sec-unitarity}}

\section{Quantum telepathy\label{sec-telepathy}}

According to the Born rule, the probability of the result $a$ of the measurement of observable $\hat A=\hat A^\dagger$
is
\be\label{Born_rule}
  p_a=\|\hat P_a\psi\|^2,
\ee
where $\psi\in\mathcal{H}$ ($\|\psi\|^2=1$) is pure state of the system before measurement,
$\hat P_a$ is orthogonal projector to eigenspace of $\hat A$
$$
  \forall \phi\in\mathcal{H}\quad \hat A\hat P_a\phi=a\hat P_a\phi.
$$

The Born rule is reliably proven on experiment.
Does there exist a non-Born experiment?
Non-Born experiment is the \emph{precise} measurement of observables with discreet spectra, which does not satisfy \eqref{Born_rule}.

\textbf{Theorem on quantum telepathy}. Let Alice holds a system component 1, which admits a Born measurement of observable $\hat A_1$.
Let Bob holds a system component 2, which admits a non-Born measurement of observable $\hat A_2$. Let the system 1+2 is described by the pure state
\begin{equation}
  |\Psi\rangle
  =\frac{|0_1\rangle|0_2\rangle+|1_1\rangle|1_2\rangle}{\sqrt2},
\label{zaput1}
\end{equation}
where $\hat A_b|a_b\rangle=a|a_b\rangle$, $\||a_b\rangle\|^2=1$.
Let Alice and Bob hold an ensemble of pairs 1+2 in state $\Psi$.
Then Alice and Bob could transmit information by measuring or by not measuring observables $\hat A_b$ at the certain time moments $t_m$. The time of information transfer could be short enough to violate the special relativity.

To transmit a bit of information from to Bob Alice at the time moment $t_a$ measures or not measures
the observable $\hat A_1$ for an ensemble of components 1.
If the measurement is performed, the state collapses
$$
  |\Psi\rangle\longrightarrow
  |0\rangle|0\rangle\textrm{~or~}|1\rangle|1\rangle.
$$
with Born probabilities $\frac12$ for each result.
Bob at the time moment $t_b>t_a$ measures (by non-Born measurement) the observable $\hat A_2$
for an ensemble of components 2 and derive the probabilities of 0 and 1.
If Bob finds Born probabilities $\frac12$, then Alice did measure $\hat A_1$.
If Bob finds non-Born probabilities $\not=\frac12$, then Alice did \textbf{not} measure $\hat A_1$.

Similarly Bob could transfer information to Alice.

So, if we believe in special relativity and formalism of description of quantum complex systems, we have to believe in Born rule.

\section{The role and place of projection postulate\label{sec-proj-need}}
  Projection postulate describes the change of quantum system state after the selective measurement of an observable.
  Projection postulate admits different modifications, so before its derivation one has to decide is it necessary to derive it.
  Is it necessary to derive a postulate, which is just a phenomenological simplified description?

  We consider schemes of precise measurement of discrete observables,
   which convert a pure quantum state to a pure quantum state.

\subsection{LL-scheme\label{subsec-LL}}
In classical textbook by L.D.~Landau and E.M.~Lifshitz \cite{Landafshic3} more general scheme is considered,
instead of projection postulate. We refer this scheme as \emph{LL-scheme}.
We consider it to demonstrate the ``flexibility'' of projection postulate, which could be easily modified.

Let $\Psi$ is the state of quantum system before a measurement.
$\Psi$ could be decomposed along eigenstates $\Psi_n$ of measured observable $\hat F$
$$
  \Psi=\sum_m c_n \Psi_n,\qquad
  \hat F\Psi_n=f_n\Psi_n,\quad
  \langle\Psi|\Psi\rangle=\langle\Psi_n|\Psi_n\rangle=1,\quad c_n=\langle\Psi_n|\Psi\rangle.
$$
  Probability $p_n$ of the measurement result $F=f_n$ is given by standard Born rule $p_n=|c_n|^2$.
  \emph{According to projection postulate} the state of quantum system after the measurement have to be just $\Psi_n$.
  According to general case of LL-scheme the state of quantum system after the measurement is
  $\varphi_n\not=\Psi_n$, $\langle\varphi_n|\varphi_n\rangle=1$.
  Moreover, set of $\varphi_n$ states could be non-orthogonal, i.e. set of $\varphi_n$ could not be describes as set of eigenstates of any Hermitian operator.

  Time duration of measurement process is not discussed in LL-scheme.

  LL-scheme is the most general description of \emph{precise} quantum selective measurement, \emph{if the initial pure state is converted into final pure state}.

  LL-scheme could be derived from combination of projection postulate and unitary evolution.
  LL-scheme measurement consists of two stages
\begin{itemize}
\item instantaneous ideal measurement, which describes by projection postulate.
  $\Psi\to\Psi_n$ with probability $p_n=|c_n|^2$.
\item Unitary evolution during short time $\delta t$ with Hamiltonian $\hat H_n$, \emph{which depends on the
result of the measurement}.
 $e^{-\frac{i}\hbar\hat H_n\,\delta t}:\Psi_n\to\varphi_n$ with probability 1.
\end{itemize}

  By selection of Hamiltonian $\hat H_n$ one can describe a measurement with arbitrary set of states $\varphi_n$.
  Moreover, one can make final state independent on the result of measurement, i.e. in this case LL-scheme describes
  the process of the preparation of a given state.

\subsection{LL-scheme and Stern-Gerlach experiment\label{subsec-SG}}
%% стрелка --- pointer

  Let us consider LL-scheme measurement in Stern-Gerlach experiment.
  During spin measurement beam of particles moves through  		
  the region of highly inhomogeneous magnetic field.
  Initial non-polarized beam is splitted into several sub-beams with definite projections of spin.
  The correlation between the coordinate of particle along the magnetic field and spin projection is created.
  This correlation allows to determine spin projection by measuring of particle coordinate.
  (One can consider a particle as set of two systems, spin and coordinate. Coordinate in Stern-Gerlach experiment
  is used as a pointer of measuring device.)
  Spin state evolution depends upon magnetic field in the region, where particle moves.
  In Stern-Gerlach experiment magnetic field is highly inhomogeneous, so different sub-beams to be exposed
  by different fields and evolve by different Hamiltonians, according to LL-scheme.
  In the original setting of Stern–Gerlach experiment spin projections in sub-beams are definite,
  so the spin evolution after the beam splitting is just different phase shift $e^{-i\omega_m\, \delta t}$, where $m$ is spin projection.
  The setting could be easily modified by additional magnets to create arbitrary magnetic fields for sub-beams.

\subsection{Projection postulate and measurements with no interaction\label{subsec-LL2}}

  In the context of LL-scheme projection postulate is just an approximation.
  One could think, that the derivation of projection postulate is senseless.
  Nevertheless, projection postulate considered to be exact in important class of measurement, the measurements with no interaction.

  Initially projection postulate is derived from the requirement that the second measurement of the same observable
  immediately after the first measurement have to produce the same result.

  This is similar to change of classical probability distribution $\rho(x)$. If a measurement indicate that $x\in[a,b]$, than
  $\rho(x)\to\chi_{[a,b]}(x)\,\rho(x)$, where $\chi_{[a,b]}$ is characteristic function of the interval $[a,b]$.
  This sort of measurement can be produced with no interaction, if device found no particle out of interval $[a,b]$.

  Projection measurement in quantum mechanics can also be considered as measurement with minimal possible interaction.
  Similarly to classical case one can (theoretically) to construct measurement to avoid interaction
  \emph{in the case of the certain result of measurement}.
  Exact time localisation and time duration of measurement in this case can be neglected.

  The other important case of projection measurement is measurement of one subsystem of complex system (of two noninteracting subsystems), then we are interested in the state of the other subsystem, which does not interact with measuring device.
  States of subsystems could be correlated by some previous interaction.
  Absence of interaction between subsystems could be provided by spatial distance.
  Isolation of subsystems in this case is guaranteed by special relativity.
  Projection postulate can be wrong with respect to whole system, but it holds for subsystem of interest, which
  does not interact with measuring device.
  Correlation between subsystems is destroyed during measurement.
  Moreover, the measured subsystem could be annihilated,
  for subsystem of interest it does not matter.
  I.e. one photon of correlated pair could be absorbed by a sensor.

  So, projection postulate is not universal, but it has to hold exactly in two important
  cases of measurement with no interaction:
\begin{itemize}
\item the selected measurement result corresponds to no interaction between system and sensor;
\item the system consists of non-interacting correlated subsystems, the projection postulate describes
 the state of subsystem, which does not interact with sensor.
\end{itemize}
  In these cases projection postulate is proven by multiple experiments.

  It makes interesting derivation (or justification) of projection postulate from the other principles
  of quantum mechanics.

\section{Derivation of projection postulate\label{sec-proj-der}}
\subsection{Two-pointer scheme\label{subsec-2a-sch}}

  We consider the process of measurement following von Neumann with two main differences:
  we consider two consequent measurements, the measured observables have discreet spectra.
  The second measurement makes reasonable the discussion on the state of the system between
  the measurements.
  Each measurement has two stages, the creation of correlation between the system and sensor pointer
  and the measurement of pointer state.
  The measurements of states of both pointers take place at the very end of the experiment.
  The state system and sensors after this moment does not considered,
  one could consider the system and sensor to be annihilated.
  So, we do not use any description of system state change, we need only Born rule for probabilities.
  The similar model was independently introduced in the paper \cite{Lesovik}.

  Let us consider the measurement of two non-commuting observables $\hat A$ and $\hat B$
  with discreet spectra.
  $a_i$ and $b_j$ are eigenvalues of  respectively.
  The corresponding orthonormal eigenvectors are
  $|a_i,k\rangle$ and $|b_j,l\rangle$. Indices $k$ and $l$ distinguish eigenvectors with same eigenvalues.
   Projectors to eigensubspaces are
$$
  \hat P_{i}=\sum_k |a_i,k\rangle\langle a_i,k|,\qquad
  \hat R_{j}=\sum_l |b_j,l\rangle\langle b_j,l|,
$$
$$
   \hat P_{i_1}\hat P_{i_2}= \hat P_{i_1}\delta_{i_1i_2},\quad
   \hat R_{j_1}\hat R_{j_2}= \hat R_{j_1}\delta_{j_1j_2},\quad
   \sum_i \hat P_{i}=\sum_j \hat R_{j}=\hat 1.
$$

  The large system, we consider consists of \emph{small system} and two \emph{pointers}.
  Small subsystem is the system, we are going to measure.
  Pointer is a microscopic part of sensor, which is correlated with small system during measurement.
  Observables $\hat A$ and $\hat B$ are acting upon small system only,
  the eigenstates $|a_i,k\rangle$ and $|b_j,l\rangle$ are states of small system.
  The states of pointer-1 and pointer-2 are designated by $\alpha$ and $\beta$ respectively.
  Basis states of pointers are $|\alpha_n\rangle$,
  $n\in\mathbb{Z}_N$ and $|\beta_m\rangle$, $m\in\mathbb{Z}_M$.
  $N$ and $M$ are large enough natural numbers\footnote{$N$
  is not less than number of different eigenvalues of $\hat A$,
  $M$ is not less than number of different eigenvalues of $\hat B$.}, or infinity.

  Let initial state of small system is $|\psi_0\rangle$.
  Initial states of pointers are $|\alpha_{0}\rangle$ and $|\beta_{0}\rangle$.
  So, initial state of large system is
$$
  |\Psi_0\rangle=|\psi_0\rangle\otimes|\alpha_{0}\rangle\otimes|\beta_{0}\rangle.
$$

  Measurement of observable $\hat A$ creates correlation of small system and pointer-1.
  State of pointer-2 (arbitrary $|\beta_{X}\rangle$) remains the same
$$
  \hat U_A: |a_i,k\rangle\otimes|\alpha_{n}\rangle\otimes|\beta_{X}\rangle\to
  |a_i,k\rangle\otimes|\alpha_{n+i}\rangle\otimes|\beta_{X}\rangle,\quad
  \langle \alpha_{i_1}|\alpha_{i_2}\rangle=\delta_{i_1i_2}.
$$
  Operator $\hat U_A$ is unitary, if sum $n+i$ is defined in $\mathbb{Z}_N$.

  Similarly, measurement of observable $\hat B$ creates correlation of small system and pointer-2.
  State of pointer-1 (arbitrary $|\alpha_{X}\rangle$) remains the same
$$
  \hat U_B: |b_j,l\rangle\otimes|\alpha_{X}\rangle\otimes|\beta_{m}\rangle\to
  |b_j,l\rangle\otimes|\alpha_{X}\rangle\otimes|\beta_{m+j}\rangle,\quad
  \langle \beta_{j_1}|\beta_{j_2}\rangle=\delta_{j_1j_2}.
$$
  Operator $\hat U_B$ is unitary, if sum $m+j$ is defined in $\mathbb{Z}_M$.

  $\hat U_A$ converts $|\Psi_0\rangle$ to
$$
 \hat U_A|\Psi_0\rangle=\hat U_A \sum_i(\hat P_{i}|\psi_0\rangle)\otimes|\alpha_{0}\rangle\otimes|\beta_{0}\rangle
 =\sum_i(\hat P_{i}|\psi_0\rangle)\otimes|\alpha_{i}\rangle\otimes|\beta_{0}\rangle.
$$

  $\hat U_B$ converts new state $\hat U_A|\Psi_0\rangle$ to
$$
 \hat U_B \hat U_A|\Psi_0\rangle=
 \hat U_B\sum_i(\hat P_{i}|\psi_0\rangle)\otimes|\alpha_{i}\rangle\otimes|\beta_{0}\rangle=
$$
$$
 =\hat U_B\sum_{ji}(\hat R_{j}\hat P_{i}|\psi_0\rangle)\otimes|\alpha_{i}\rangle\otimes|\beta_{0}\rangle=
$$
$$
 =\sum_{ji}(\hat R_{j}\hat P_{i}|\psi_0\rangle)\otimes|\alpha_{i}\rangle\otimes|\beta_{j}\rangle
$$
  Each term of the superposition corresponds to a certain position of pointers.
  Pointer-1 indicates $A=a_i$, pointer-2 indicates $B=b_j$.
  Probability of this result is
$$
p_{ij}=\|\hat R_{j}\hat P_{i}|\psi_0\rangle\|^2=
\langle\psi_0|\hat P_{i}\hat R_{j}^2\hat P_{i}|\psi_0\rangle
=\langle\psi_0|\hat P_{i}\hat R_{j}\hat P_{i}|\psi_0\rangle.
$$
$$
  p_{ij}=\langle\psi_{Ai}|\hat R_{j}|\psi_{A_i}\rangle.
$$
  Here $|\psi_{Ai}\rangle=\hat P_{i}|\psi_0\rangle$.

  According to this description observer measures states of both pointers simultaneously at the end of experiment.
  So, measurement of two non-commuting observables $\hat A$ and $\hat B$ is reduced to measurement
  of two commuting observables, the states of two pointers.

  Let us compare this description with description, which uses projection postulate.
  The probability of pointer-1 readings $A=a_i$ is
$$
  p_i=\sum_{j}p_{ij}=\sum_j\langle\psi_0|\hat P_{i}\hat R_{j}\hat P_{i}|\psi_0\rangle
  =\langle\psi_0|\hat P_{i}\underbrace{\left(\sum_j\hat R_{j}\right)}_{\hat 1}\hat P_{i}|\psi_0\rangle=
$$
$$
  =\langle\psi_0|\hat P_{i}|\psi_0\rangle=\langle\psi_{Ai}|\psi_{Ai}\rangle.
$$

  Conditional probability that the pointer-2 indicates $B=b_j$, under the condition that the pointer-2
  indicates $A=a_i$ is
$$
  p_{j|i}=\frac{p_{ij}}{p_i}=\frac{\langle\psi_{Ai}|\hat R_{j}|\psi_{Ai}\rangle}{\langle\psi_{Ai}|\psi_{Ai}\rangle}.
$$

  Conditional probability $p_{j|i}$ is equal to probability to find measurement result $B=b_j$
  for the state $|\psi_{Ai}\rangle=\hat P_{i}|\psi_0\rangle$, which is derived from initial state by
  projection postulate.

  Projection postulate could be interpreted as transition to conditional probability amplitudes.

  The derivation of projection postulate is model dependent (see section \ref{subsec-LL}).

\subsection{One-pointer scheme\label{subsec-1a-sch}}

  We can simplify the two-pointer scheme by removing the pointer-2 and action of operator $\hat U_B$.
  Instead of pointer-2 one can use the small system itself.
  The one-pointer scheme instead of measuring of states of two pointer the finishing measurement is measurement of state of pointer-1 and direct measurement of observable $\hat B$ for small system.

$$
  |\Psi_0'\rangle=|\psi_0\rangle\otimes|\alpha_{0}\rangle.
$$
$$
  \hat U_A: |\psi_{Aik}\rangle\otimes|\alpha_{n}\rangle\to
  |\psi_{Aik}\rangle\otimes|\alpha_{n+i}\rangle.
$$
$$
 \hat U_A|\Psi_0'\rangle=
 \sum_i(\hat P_{i}|\psi_0\rangle)\otimes|\alpha_{i}\rangle
 =\sum_{ji}(\hat R_{j}\hat P_{i}|\psi_0\rangle)\otimes|\alpha_{i}\rangle
$$
  In the last formula term $(\hat R_{j}\hat P_{i}|\psi_0\rangle)\otimes|\alpha_{i}\rangle$
  corresponds to the measurement result $A=a_i$, $B=b_j$.

  Probabilities $p_{ij}$ and  conditional probabilities $p_{j|i}$ are the same as in two-pointer scheme.

\subsection{One-pointer scheme and EPR-experiment\label{subsec-1a-sch-EPR}}

  Gedankenexperiment by Einstein, Podolsky and Rosen (EPR-experiment) is based upon the idea of reducing
  measurement of non-commutative observables to measurement of commuting observables related to different
  subsystems.
  This idea coincide with idea of two-pointer and one-pointer schemes above.

  Let us demonstrate that EPS-experiment in Bohm modification \cite{Bohm} (see p. 661 The Paradox of Einstein, Rosen, and Podolsky) corresponds to one-pointer scheme, described above.

  Let small system is q-bit with the initial state $|\psi_0\rangle=\frac{|\uparrow\rangle-|\downarrow\rangle}{\sqrt2}$.
  Pointer is q-bit with initial state $|\uparrow\rangle$.
$$
  |\Psi_0'\rangle=\frac1{\sqrt2}(|\uparrow\rangle-|\downarrow\rangle)|\uparrow\rangle.
$$
  The first observable is $\hat A=\hat \sigma_z$.
  Operator $\hat U_A$ is ``conditional not'',
  which inverse the state of pointer if small system is in the state $|\uparrow\rangle$
$$
  \hat U_A: |\uparrow\rangle|x\rangle\to
  |\uparrow\rangle(\underbrace{\hat\sigma_x|x\rangle}_{|\textrm{not~} x\rangle}),\quad
  |\downarrow\rangle|x\rangle\to|\downarrow\rangle|x\rangle.
$$
  Acting by $\hat U_A$ on $|\Psi_0'\rangle$ we get EPR-state of small system and pointer
$$
  \hat U_A|\Psi_0'\rangle=\frac1{\sqrt2}(|\uparrow\rangle|\downarrow\rangle-|\downarrow\rangle|\uparrow\rangle).
$$
  Finally $\hat \sigma_z$ is measured for second q-bit (pointer) and arbitrary 1-bit operator $\hat B$
  is measured for first q-bit (small system).

\section{Conclusion\label{sec-concl}}

Born rule for probabilities is derived from no-telepathy condition.
This derivation uses weakened version of projection postulate, for complex correlated system without interaction
during measurement process.

Instead of attempts to derive von Neumann projection postulate from unitary evolutions, we derive projection postulate
from Born rules and correlations of measurement results.

\subsection*{Acknowledgement}

We thank I.V.~Volovich, Yu.M.~Belousov, S.N.~Filippov, G.B.~Lesovik, V.I.~Man'ko, N.N.~Shamarov, O.G.~Smolyanov and paricipants of Seminar on Quantum Physics and Quantum Information in MIPT and seminar of O.G.~Smolyanov in MSU for discussion.

\end{document}